\documentclass[aps,pra,amsmath,twocolumn,superscriptaddress]{revtex4-1}
\usepackage{graphicx}
\graphicspath{ {images/} }
\usepackage{dcolumn}
\usepackage{bm}
\usepackage{color,soul}

\begin{document}

\preprint{APS/123-QED}

	\author{S. Flannigan} \affiliation{Department  of  Physics  $\&$  SUPA,  University  of  Strathclyde,  Glasgow  G4  0NG,  United  Kingdom.}
	\author{L. Madail}\affiliation{Department of Physics $\&$ I3N, University of Aveiro, 3810-193 Aveiro, Portugal.}
	\author{R. G. Dias} \affiliation{Department of Physics $\&$ I3N, University of Aveiro, 3810-193 Aveiro, Portugal.}
	\author{A. J. Daley} \affiliation{Department  of  Physics  $\&$  SUPA,  University  of  Strathclyde,  Glasgow  G4  0NG,  United  Kingdom.}


\title{Hubbard models and state preparation in an optical Lieb lattice}

\date{\today}

\begin{abstract}
Inspired by the growing interest in probing many-body phases in novel two-dimensional lattice geometries we investigate the properties of cold atoms as they could be observed in an optical Lieb lattice. We begin by computing Wannier functions localised at individual sites for a realistic experimental setup, and determining coefficients for a Hubbard-like model. Based on this, we show how experiments could probe the robustness of edge states in a Lieb lattice with diagonal boundary conditions to the effects of interactions and realise strongly correlated many-body phases in this geometry.
We then generalise this to interacting particles in a half-filled 1D Lieb ladder, where excitations are dominated by flat band states. We show that for strong attractive interactions, pair correlations are enhanced even when there is strong mixing with the Dirac cone. 
These findings in 1D raise interesting questions about the phases in the full 2D Lieb lattice which we show can be explored in current experiments. 
\end{abstract}

\maketitle


\section{Introduction}
In recent years, there has been much progress in realising and exploring novel lattice geometries using ultra-cold atoms in optical lattices ~\cite{Bloch:2012aa,RevModPhys.80.885}, and especially lattices with topological properties
~\cite{RevModPhys.91.015005,PhysRevLett.110.185301,PhysRevLett.112.156801,PhysRevB.91.245135,Aidelsburger:2018ab,PhysRevB.97.201115,PhysRevLett.119.180402,PhysRevA.93.033605,PhysRevB.92.115120,PhysRevA.89.033632,PhysRevLett.117.163001}. While a lot of work has focused on few-particle properties, there is increasing interest in the interplay between topological band structures and strong interactions. For example, recent studies have explored how lattice topology can enhance superfluid phases with cold gases ~\cite{PhysRevLett.117.163001,PhysRevB.82.184502,SF_Cr} or potentially superconducting properties in solid-state systems ~\cite{PhysRevLett.117.045303,PhysRevLett.124.207006,PT_Top}, as a result of the interplay between interactions and flat energy bands. 

\begin{figure}[t!]
{\includegraphics[width = 8.5cm]{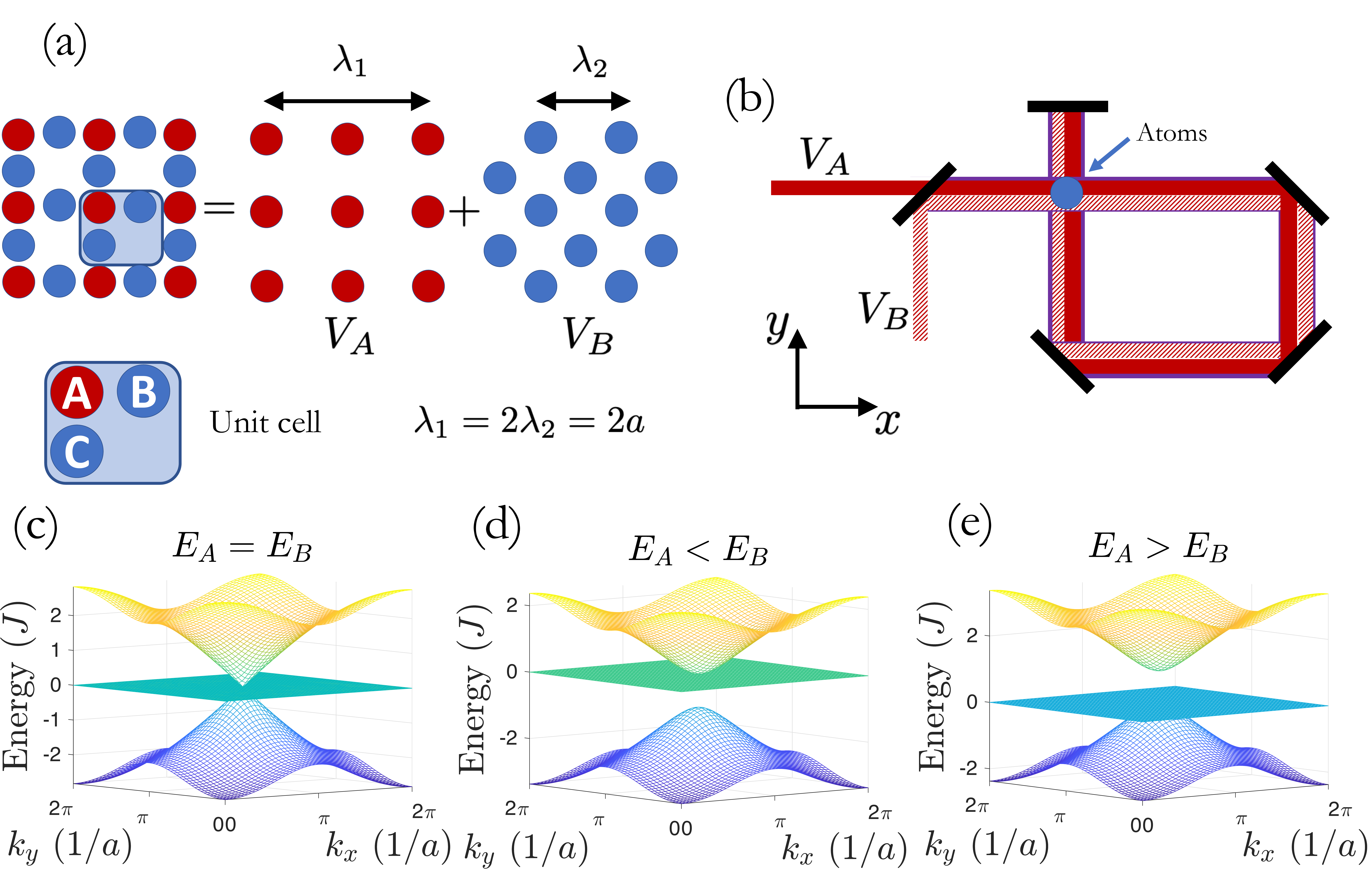}}\\ 
\caption{(a) Creation of an optical Lieb lattice by superimposing two optical potentials, $\{ V_A,V_B \}$ given by Eq.~\ref{equpot}. (b) Laser configuration for each optical potential, following the scheme for realisation of the Lieb lattice proposed by Di Liberto et al. in Ref.~\cite{PhysRevLett.117.163001}. The incoming solid red laser (for $V_A$) is polarised along the $y$-direction, while the dashed patterned laser (for $V_B$) is polarised in the direction orthogonal to the $x$-$y$ plane. (c-e) Single particle energy band structure for a tight-binding model on the Lieb lattice for different ratios of the onsite energies, $\{ E_A, E_B, E_C \}$, for each site of the unit cell ($E_C=E_B$ for all), given in units of the nearest-neighbour tunnelling rate, $J$, and the unit cell spacing, $a$.}
\label{Lieb_Lattice}
\end{figure}

In this work, we consider the potential experimental realisation of interacting particles in an optical Lieb lattice, as shown in Fig.~\ref{Lieb_Lattice}(a). This can be realised in experiments either by using a quantum gas microscope, or following schemes such as that proposed by Di Liberto et al. in Ref.~\cite{PhysRevLett.117.163001} and sketched in Fig.~\ref{Lieb_Lattice}(b). We begin by numerically determining localised Wannier functions on individual lattice sites and calculating coefficients for a Hubbard model based on experimentally accessible parameters. We then consider interacting particles in these systems in two example cases of weak and strong interactions. We first probe the robustness of edge and corner states in a Lieb lattice with diagonal boundary conditions~\cite{PhysRevB.100.125123} when weak interactions are introduced, and then we investigate the many-body phases manifested in a half-filled 1D Lieb ladder for strong attractive interactions. 
These calculations provide a potential roadmap for experimental investigation of many-body phases for cold atoms in a Lieb lattice.

As depicted in Fig.~\ref{Lieb_Lattice}(c), the Lieb lattice has a flat energy band that touches a dispersive band at a single $k$-point, which becomes a Dirac cone~\cite{Gardenier:2020aa} for uniform onsite energies. Dispersionless (or flat) single particle energy bands arise through a destructive interference effect of the single particle wavefunction, \cite{doi:10.1080/23746149.2018.1473052,Liu_2014,PhysRevLett.81.5888,Mukherjee:15,PhysRevB.101.224514,Hub_FB,PT_FB_Lieb}, leading to an infinite energy degeneracy and means that conventional single-particle tunnelling is suppressed for particles in superpositions of states belonging to this flat band. It has been shown however, that interactions allow the atoms to form pairs that are mobile~\cite{PhysRevB.98.220511}. For bosons this can push the ground state of the system towards forming a pair condensate if the lowest energy band is flat~\cite{PhysRevA.88.063613,PhysRevB.88.220510,PhysRevLett.108.045306,PhysRevB.92.195149,SF_Cr} and for fermions can result in an enhanced critical temperature for superconductivity~\cite{PhysRevB.83.220503,Heikkila:2011aa}, with phases that have large pair correlations~\cite{PhysRevLett.117.045303,PhysRevB.98.094513}. But there are additional open questions here regarding the corrections to previous theoretical predictions, particularly when the interaction strengths are strong enough to mix states in different bands.

In order to analyse the effects of strongly interacting systems in novel band structures it is necessary to derive effective models which can describe the dynamics induced in these systems. While it is now routine to realise Hubbard models in simple periodic potentials, up to well controlled approximations~\cite{Jaksch:2005aa,RevModPhys.80.885,PhysRevLett.81.3108,Muller2012}, it can still be a challenging task to derive the coefficients in these models for more complicated lattices. In order to allow future experiments to more easily explore these questions in the Lieb lattice, we first derive experimentally realistic values for the coefficients in a Hubbard model as a function of the laser intensity through calculating the single particle Wannier functions localised on each site~\cite{PhysRevA.87.043613}. We then show that one could utilise this knowledge in probing the effects of interactions on the edge states that are manifested in a rotated Lieb lattice, where we find that for sufficiently dilute and weakly interacting atoms that they are robust to these effects. 
We also consider the many-body fermion case at half-filling where we show that pair correlations appear quasi-long range. These strongly dominate over exponentially suppressed single particle correlations and are sufficiently large to be resolved experimentally. Furthermore, we show that the pair correlations become further enhanced when there is greater mixing between states in the flat band and in the higher dispersive bands due to increasing the interaction strength. Finally, in order to be able to realise these phases experimentally we describe and demonstrate an experimentally feasible adiabatic ramp process for preparing these ground states.

The remainder of this article is organised as follows. In section \ref{sec:hubbard} we compute Wannier functions and determine Hubbard model parameters from potential experimental parameters with cold atoms. In section \ref{sec:edge}, we then discuss geometrically protected edge states, and the effects of interactions on mixing of those states, before discussing strongly interacting systems of fermions at half-filling in a Lieb ladder in section \ref{sec:fermions}. We discuss how these states might be prepared in experiments in section \ref{sec:expt} before giving a summary and outlook in section \ref{sec:conclusions}.

\section{Localised Wannier functions and derivation of  Hubbard coefficients}
\label{sec:hubbard}
In the Lieb lattice for the case where every site is degenerate the system manifests a three band touching point, with a Dirac cone that crosses the flat band, but by tuning the ratio of the onsite energies it is possible to realise different energy gaps, see Fig.~\ref{Lieb_Lattice}. Here we consider producing the Lieb lattice experimentally by superimposing two optical potentials~\cite{PhysRevLett.117.163001},
\begin{equation}\label{equpot}
 \begin{split}
V_{A}(x,y) &=-V_{1}[\cos^{2}(k_{1}x)+\cos^{2}(k_{1}y)] \\
V_B(x,y) &= -\frac{V_2}{4} [\cos(k_{2}x) - \cos(k_{2}y)]^2,
 \end{split}
\end{equation}
where the wavevector $k_{i} = {2\pi}/{\lambda_{i}}$, and $\lambda_{1} \approx 2\lambda_{2}$. In Fig.~\ref{Lieb_Lattice}(a) we illustrate the direction of the beams and the values required for the wave-vectors. Usually, $V_1, V_2$ are expressed in units of the recoil energy, $E_{Ri}= \hbar^2 k_i^2/2m$, where $m$ is the mass of the atoms. The above potential can be produced experimentally using the configuration shown in Fig.~\ref{Lieb_Lattice}(b) which follows from the proposal in Ref.~\cite{PhysRevLett.117.163001}. For $V_A$ (red) we use laser light that is polarised in the plane of the lattice and for $V_B$ (blue) we use a polarisation that is orthogonal to the plane. We note that this geometry has recently been realised by instead superimposing three optical potentials~\cite{Taiee1500854,PhysRevLett.118.175301}. The main differences between this approach and the one that we have considered, is that it requires three optical potentials, instead of only two in this scheme, but has the advantage of not requiring orthogonal polarisations between the beams. Also note that these lattices could alternatively be produced directly through spatial light modulation in a quantum gas microscope \cite{Mazurenko:2017aa,Liang:10}.

By tuning the ratio $R=V_{2}/V_{1}$ (achieved experimentally by varying the intensities of the two superimposed lasers) around $1$, the relative depth of the trap at the $A$ sites and the $B,C$ sites can be varied, which then modifies the single particle energy band structure, see Fig.~\ref{Lieb_Lattice}(c-e). Under well controlled approximations it has been shown that atoms confined in an optical lattice can be described with a Hubbard model~\cite{Bloch:2012aa,RevModPhys.80.885,Jaksch:2005aa} for sufficiently large potential depths. We will see below that for the Lieb lattice we require laser intensities of around $V_1, V_2 \sim 40 E_{R2}$ so that the next nearest-neighbour tunnelling rates are below $1\%$ of the nearest-neighbour components. In this case the Hubbard model on this geometry for two species fermions takes the form ($\hbar \equiv 1$)
\begin{equation}\label{Ham_Lieb}
\begin{split}
H = & \sum_{n,\sigma} E_A a_{\sigma,A,n}^{\dagger} a_{\sigma,A,n} + E_B a_{\sigma,B,n}^{\dagger} a_{\sigma,B,n} + E_C a_{\sigma,C,n}^{\dagger} a_{\sigma,C,n} \\
& - J \left( a_{\sigma,A,n}^{\dagger} a_{\sigma,B,n} + a_{\sigma,A,n}^{\dagger} a_{\sigma,C,n} + a_{\sigma,B,n}^{\dagger} a_{\sigma,A,n-\vec{x}} \right. \\
& \left. + a_{\sigma,C,n}^{\dagger} a_{\sigma,A,n+\vec{y}} + h.c \right) \\
& + U_A \hat{n}_{\sigma,A,n} \hat{n}_{\overline{\sigma},A,n} + U_B \hat{n}_{\sigma,B,n} \hat{n}_{\overline{\sigma},B,n} + U_C \hat{n}_{\sigma,C,n} \hat{n}_{\overline{\sigma},C,n} ,
\end{split}
\end{equation}
\noindent where $a_{\sigma,\kappa,n}^{\dagger}$ creates an atom at the unit cell $n$ on site $\kappa$ and $\hat{n}_{\sigma,\kappa,n} = a_{\sigma,\kappa,n}^{\dagger}a_{\sigma,\kappa,n}$. Additionally, $\vec{x}$ ($\vec{y}$) denotes the translation of one unit cell in the $x$($y$)-direction. For fermions, $\overline \sigma$ denotes the opposite spin species. 

\begin{figure}[t!]
\includegraphics[width = 8.5cm]{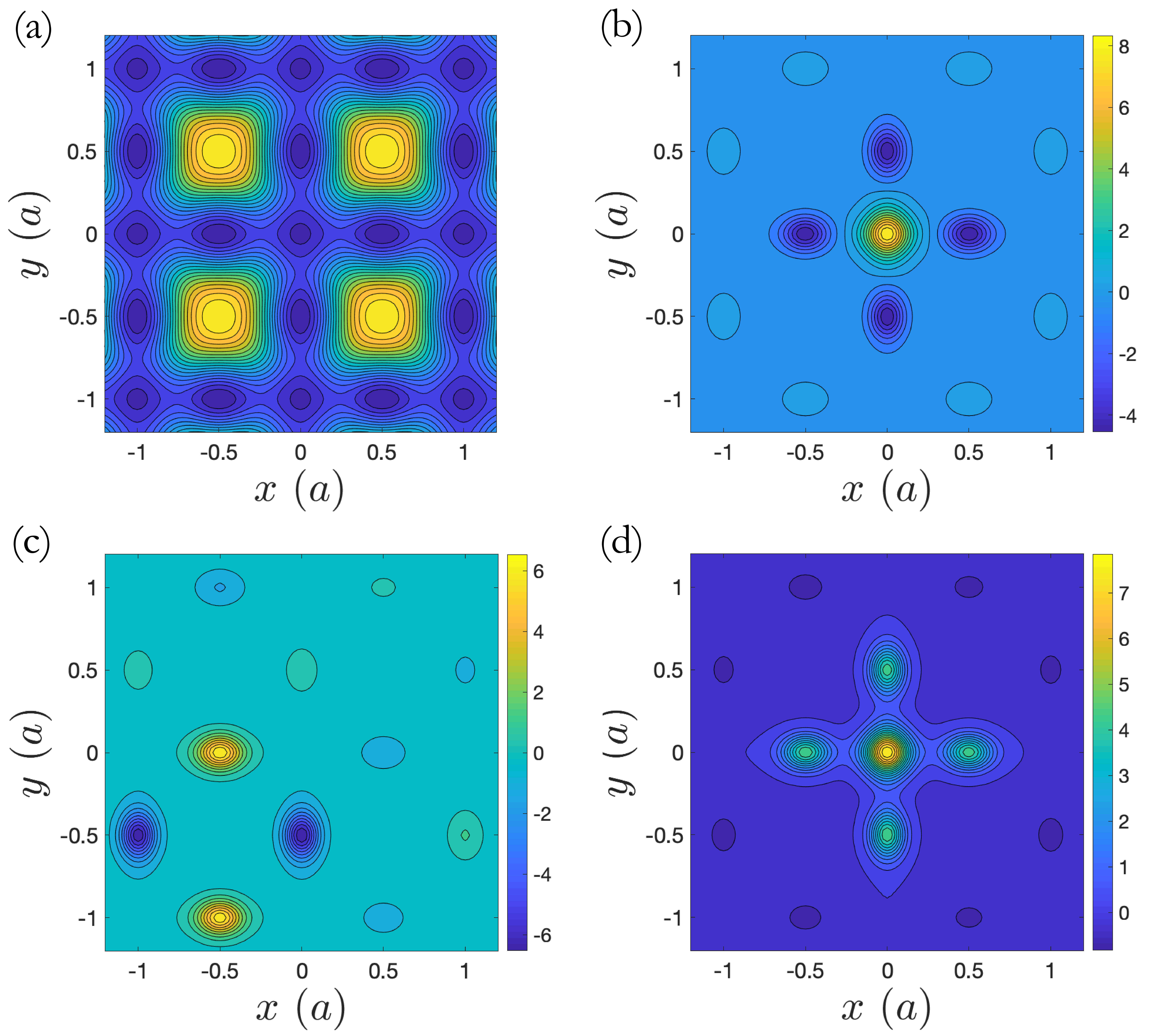}
\centering
\caption{(a) Optical potential for the Lieb lattice produced using Eq.~\ref{equpot} with $V_{1} = 34.5 E_{R2}$ and $R=V_{2}/V_{1}=1.07$. The Wannier functions associated with each energy band for a single particle in the optical potential: (b) higher band; (c) flat band; (d) lower band.}
\label{Wann_Lieb}
\end{figure}

The coefficients in this model can be calculated through the single particle Wannier functions, which can be found by first diagonalising the single particle Hamiltonian, to obtain the Bloch functions, $\phi_{k}^{m}(\vec{r})$, associated with each band $m$. We can then calculate the Wannier functions,
\begin{equation}\label{equation_Wan_Vec}
w_{j}^{n}(\vec{r}) = \frac{1}{2\pi}\int dk U^k_{n,m} \phi_{k}^{m}(\vec{r})e^{-i\vec{k}\cdot\vec{R_{j}}},
\end{equation}
where $U^k_{n,m}$ is a $k$-dependent $3\times3$ unitary matrix. For the case where each $U^k$ is diagonal and we find the appropriate choice for the phase factors then we can obtain unique and real Wannier functions that are exponentially localised to a single unit cell and each represent particles in an individual energy band~\cite{PhysRev.115.809}. This phase is chosen so that the Bloch functions are as smoothly varying as possible around the Brillouin zone. In this way, the Wannier functions for the Lieb lattice can be found as shown in Fig.~\ref{Wann_Lieb}. 

However, calculating the terms in the Hubbard model using these ordinary Wannier functions would result in a model with many long range tunnelling and interaction processes. It is then advantageous to calculate a new set of basis states which themselves are localised on individual sites, and will allow for the realisation of a Hubbard model of the form of Eq.~\ref{Ham_Lieb} with only nearest-neighbour processes. In general, this can be a non-trivial task particularly when the lattice geometry has a complicated spatial structure and connectivity. To this end, we use the method presented in Ref.~\cite{PhysRevA.87.043613} which finds the optimal superposition of the ordinary Wannier functions, by iteratively optimising each $U^k_{n,m}$, such that we obtain a new set of orthogonal functions, but which are much more localised. The effect of the off-diagonal elements of $U^k_{n,m}$ is to mix the Bloch functions of each energy band, thus the new basis set corresponds to particles localised on individual lattice sites, but which exist in a superposition of the three energy bands. 

While we encountered no technical problems in applying this iterative method in this case, it is possible for the algorithm to find local minima giving rise to a sub-optimal Wanner basis. Note that an alternative method for calculating the maximally localised onsite Wannier function has been proposed in Ref.~\cite{PhysRevLett.111.185307} which allows for a non-iterative way of calculating the band mixing matrix $U^k_{n,m}$ in Eq.~\ref{equation_Wan_Vec}, which may lead to advantages for certain systems. 

\begin{figure}[t!]
{\includegraphics[width = 8.5cm]{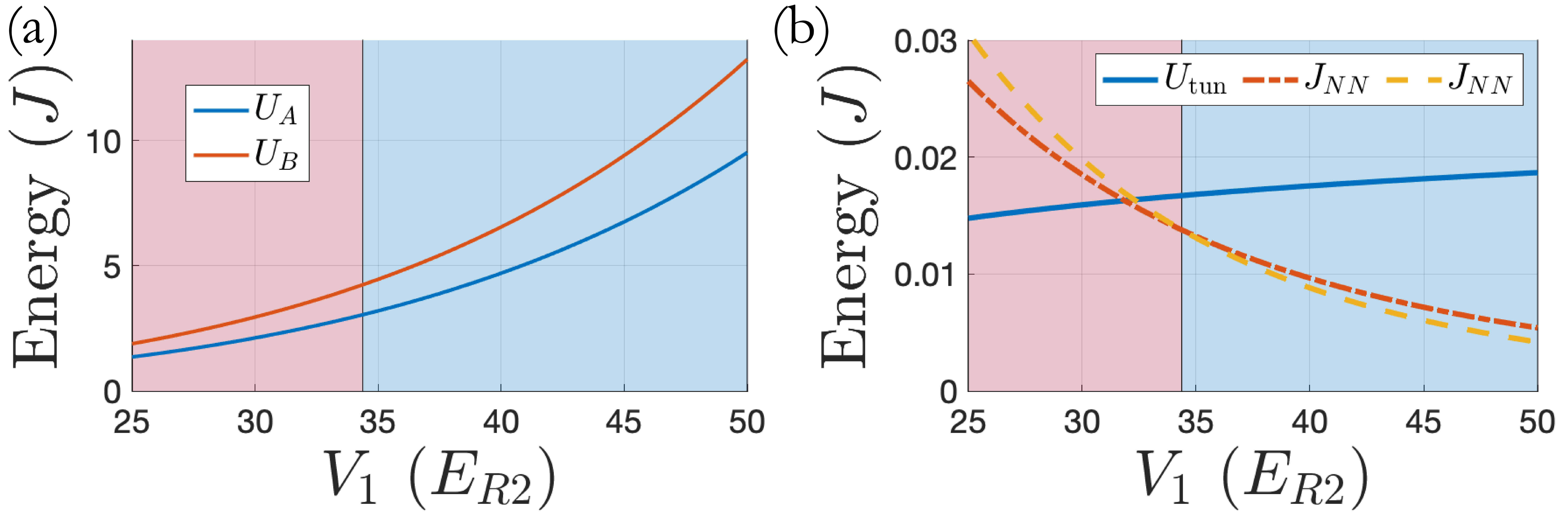}}\\ 
\caption{Local Hubbard coefficients produced from Eq.~\ref{equpot} as a function of the laser intensity, calculated with a maximally localised Wannier basis for each site. In the red zone the flat band is degenerate with the higher dispersive band (see Fig.~\ref{Lieb_Lattice}(d)), in the blue zone it is degenerate with the lower energy band (see Fig.~\ref{Lieb_Lattice}(e)) and at the boundary the band structure manifests the three band touching point and Dirac cone (see Fig.~\ref{Lieb_Lattice}(c)). (a) $U_A$ ($U_B$) are the onsite interaction strengths for the $A$ ($B$) sites of the unit cell, where $U_C=U_B$. (b) $J_{NN}$ are the next-nearest-neighbour tunnelling rates between $A$ ($B$) sites in different unit cells, red (yellow) and $U_{\rm tun}$ are the dominant interaction strengths between nearest-neighbour sites. We have used $V_2=1.07V_1$ and wavelength $\lambda_1=1064~{\rm nm}$, with a scattering length $a_s=50a_0$ and have considered $^{39}K$ atoms confined in a harmonic trap in the z-direction with $\omega_z=2\pi\times 20~{\rm kHz}$. For all values the nearest-neighbour tunnelling is on the order of $J\sim 100~{\rm Hz}$. }
\label{Lieb_Coeffs}
\end{figure}

Our calculated maximally localised onsite basis is most convenient for deriving the coefficients of a Hubbard model because we can use them to obtain the tunnelling amplitudes between nearest-neighbouring sites and the onsite energy and interaction coefficients as a function of the optical lattice laser potential values which is important for experimentally realising theoretical predictions that require knowledge of the interaction strength and are paramount for being able to reliably explore the physics of many-body systems in complex geometries. These coefficients are shown in Fig.~\ref{Lieb_Coeffs} for the regimes where the tight-binding spectrum is well approximated, where we can see that the onsite interaction strengths can be tuned to values close to the tunnelling rates for moderate scattering lengths. Comparing these maximally localised Wannier functions to those calculated in Ref.~\cite{Taiee1500854} for the alternative experimental realisation, we find very similar results. In particular, we also find that the Wannier states centred on the B/C sites to be more localised than those centred on the A sites, which we believe will result in the same asymmetry in the onsite interaction values found in this description. 

Note that compared to a conventional square lattice, which requires intensities of $V_0 > 5 E_R$ in order to accurately approximate a tight-binding model, we find here that we must use a larger laser intensity,  $V_0 > 25 E_R$ in order to reproduce a spectrum similar to the tight-binding case in Fig.~\ref{Lieb_Lattice}(c-e). However, as can be seen from Fig.~\ref{Wann_Lieb}(a) the potential is greatly dominated by the large peaks which creates the distinct pattern of the Lieb lattice and the potential barriers between nearest-neighbouring sites are much lower, giving values for the nearest-neighbour tunnelling rates on the order of $J\sim 100~{\rm Hz}$ for all parameters considered in Fig.~\ref{Lieb_Coeffs}. This means that while we require a higher laser power to produce the tight-binding spectrum compared to conventional lattices, the dynamics manifested in this system will be at a similar rate. 

We are also able to calculate corrections to a standard Hubbard model, such as longer range off site processes. In Fig.~\ref{Lieb_Coeffs}(b) we plot the next-nearest-neighbour single particle tunnelling rate and the largest nearest-neighbour density assisted tunnelling coefficient, which is explicitly given by
\begin{equation}\label{equ_Dens}
U_{\rm tun} \propto g \int d\vec{r}~w^*_A(\vec{r}) |w_B(\vec{r})|^2 w_B(\vec{r}),
\end{equation}
where $g = 4\pi\hbar^2 a_s/2m$, with $a_s$ the characteristic scattering length and $w_A(\vec{r})$, $w_B(\vec{r})$ are the Wannier functions centred on an A site and B site respectively. We can see that these values for all of these correction terms are on the order of $1\%$ for the potential values illustrated here. Additionally, the increasing behaviour of the density assisted interaction term as the potential strength is increased arises from the non-trivial interplay between the two optical potentials. As such, increasing the potential barrier height has a great effect on the large potential peaks producing the Lieb lattice pattern, which increases the confinement of the B/C sites, in the directions orthogonal to the tunnelling between these sites and the nearest A sites (see the elliptical shape of these sites in Fig~\ref{Wann_Lieb}(a)). This means that in Eq.~\ref{equ_Dens} the weight of the onsite terms $|w_B(\vec{r})|^2 w_B(\vec{r})$ increases and will lead to an increase in $U_{\rm tun} $. In a conventional lattice, this effect is usually counteracted by a much smaller overlap between nearest-neighbour Wannier functions which then results in these tunnelling rates decreasing as the potential is increased, however for the case considered here, the barrier height between the A and B sites, is only weakly affected by an increase in the potential and so the overlap between $w_A(\vec{r})$ and $w_B(\vec{r})$ does not decrease rapidly enough to counteract the effects of the increased confinement of the B sites. Thus resulting in a density assisted tunnelling rate that slowly grows with increasing potential depth for the regime considered here.

The most dominant effect of the next nearest-neighbour tunnelling term is to introduce a small curvature and bandwidth for the flat energy band, however, this is also much smaller than the dominant nearest-neighbour tunnelling rates ($< 1\%$) indicating that these corrections will have a negligible effect on the dynamics as they are significantly dominated by the conventional Hubbard terms. 

\section{Geometrically protected edge states}
\label{sec:edge}
\begin{figure}[t!]
{\includegraphics[width = 8.5cm]{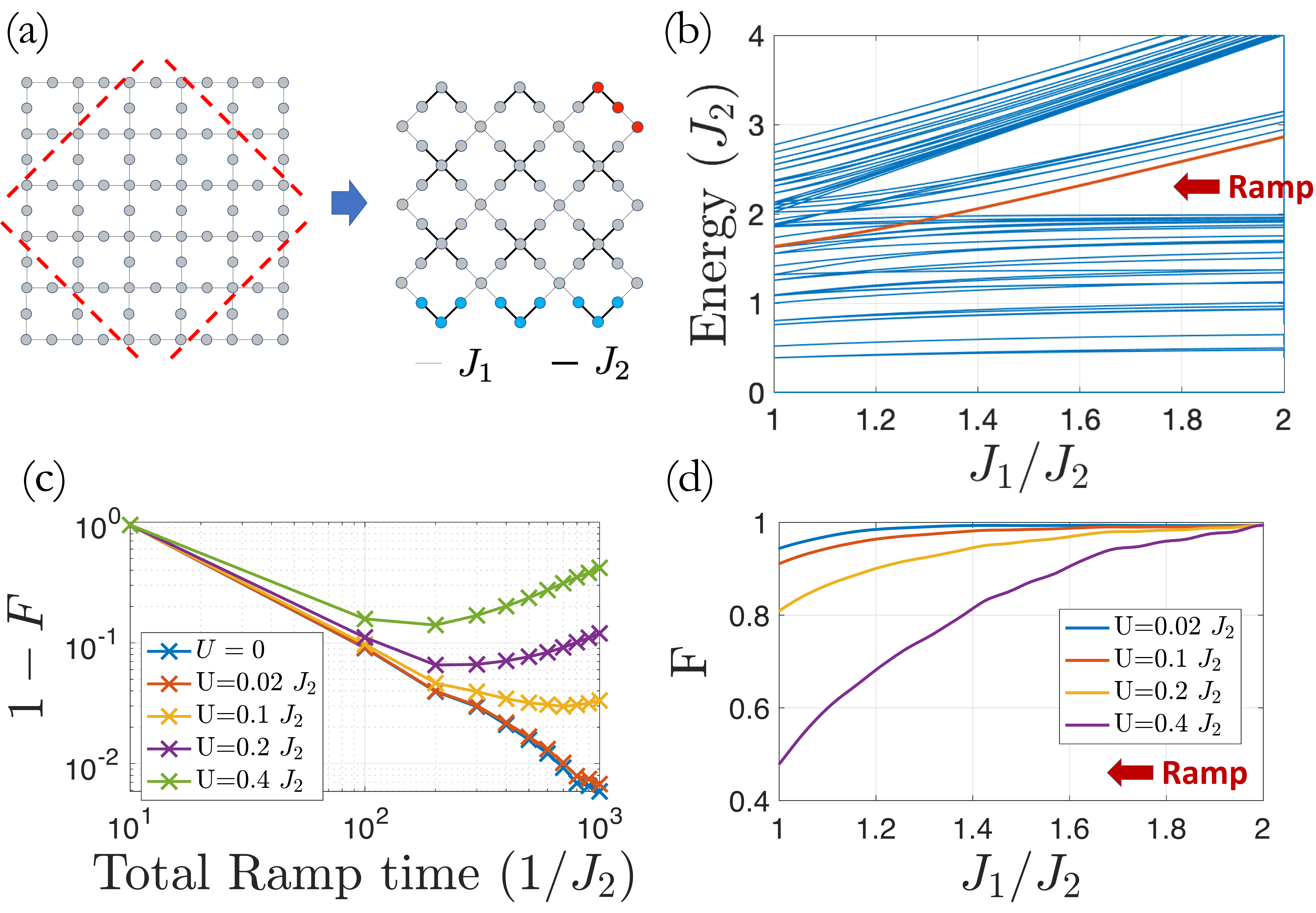}}
\caption{(a) The creation of the rotated lattice from the two-dimensional Lieb model by applying potential boundaries along the red dashed lines. We also include the spatial distribution of a particular compact edge state (blue) which exist for all $J_1/J_2$ and the corner states (red) which exist at $J_1=J_2$. (b) Positive part of the symmetric energy spectrum as a function of $J_1/J_2$ for a system size of $5\times5$ plaquettes. The red line shows the adiabatic path of the edge state (beginning at $J_1=2J_2$) into the corner state at $J_1=J_2$. (c) Time-dependent adiabatic transformation of an edge eigenstate into a corner eigenstate for the square rotated Lieb lattice. Error in the projection, $F$, of the final produced state onto all corner eigenstates of the final (non-interacting) Hamiltonian ($J_1/J_2 = 1$) as a function of the total ramp time and for different interaction strengths with $10\times 10$ plaquettes. (d) Time dependent projection onto the state with both particles in the single particle edge eigenstate highlighted in red on the energy spectrum shown in (b) for different interaction strengths. Note that this state corresponds to the corner state when $J_1=J_2$. $J_2T=100$ and $5\times 5$ plaquettes.}
\label{Rot_Lieb}
\end{figure}

A first application of studying interactions in this setup is to understand how single-particle properties are affected by interactions, and to what extent these are modified in experiments with unavoidable weak interactions. A good example of interesting single-particle phenomena for the Lieb lattice occurs when we consider a rotated Lieb lattice with a boundary as shown in Fig.~\ref{Rot_Lieb}(a), which would require additional optical potentials to act as a hard boundary. 
This model has edge states for all values of dimerisation $J_1/J_2$ and corner states for the $C_4$ symmetric regime at $J_1=J_2$. In this system, the corner states are similar to edge states in the sense they are localised at the systems boundary and have an exponentially decaying amplitude in the bulk towards the centre of the lattice. The main differences are that the corner states also have exponential suppression for amplitudes towards the middle of the edges and so are localised to the corners. Additionally, they do not exist in an energy gap, and are instead degenerate with the bulk bands.
The subset of edge and corner states that we will consider here, are those that are produced through a geometrical frustration effect similar to the mechanism that gives rise to the bulk flat band states, where as the particle attempts to move away from the edge (or corner) the various tunnelling components from the eigenstate onto the nearest bulk state interfere destructively, suppressing transport and creating localised states. We illustrated in Ref.~\cite{PhysRevB.100.125123} by considering periodic boundary conditions in the $y$-direction, that these geometrically enhanced edge states exist at the $ak_y=\pi$ point in the Brillouin zone.

Also, in Ref.~\cite{PhysRevB.100.125123} we showed that for non-interacting particles, these edge states can be prepared by an adiabatic ramping protocol, where we simulated a time-dependent variation of the value of the tunnelling dimerisation $J_1/J_2$. In Fig.~\ref{Rot_Lieb}(b) we illustrate the energy spectrum throughout this time-dependent process, where we begin in an edge state at $J_1/J_2=2$, with similar spatial structure to the one shown in blue in Fig.~\ref{Rot_Lieb}(a) and linearly ramp this ratio to $J_1=J_2$ in a set ramp time. In Ref.~\cite{PhysRevB.100.125123} we found that even though the time-dependent state crosses through many bulk energy states (blue lines) that there were no avoided crossings and we were able to prepare the final corner state, indicating that the geometrical frustration mechanism suppresses transport away from the edge and protects mixing with the bulk states even in the presence of time-dependent perturbations. 

Here, we can consider how interactions affect this protocol, and how small the interactions should be so as to ensure that the final state can be reached with high-fidelity in this ramping protocol. We employ the same ramp process, but now with two (bosonic and same species) atoms in the initial edge state that can interact with an onsite interaction strength, $U$. In Fig.~\ref{Rot_Lieb}(c) we plot the final projection of our produced state onto the corner eigenstates of the final Hamiltonian where $U=0$ for a range of ramp times and onsite interaction strengths. As expected it is apparent that as we increase the interaction strength that the final fidelity decreases, indicating that the interactions can cause mixing between the edge states and the bulk states somewhat breaking the geometrical frustration. It is also clear that beyond weak interactions $U>0.1J_2$, increasing the ramp time beyond some value dependent on $U$ leads to lower fidelities. This is because we calculate the fidelities with respect to the $U=0$ eigenstates, so this feature indicates that the eigenstates for $U\neq0$ have a decreasing overlap with the non-interacting states as $U$ is increased. In some cases, this can then result in a better preparation of the $U=0$ corner states for non-adiabatic processes at shorter ramp times.
We further probe this in Fig.~\ref{Rot_Lieb}(d) where we calculate the time-dependent projection throughout a single ramp ($J_2T=100$) of the current state onto the product state where both particles are in the single particle eigenstate highlighted on the energy spectrum in red in Fig.~\ref{Rot_Lieb}(b). Comparing different strengths of interactions we illustrate that for weak interactions ($U<0.2J$), the majority of the mixing occurs when the edge state crosses the bulk states for $J_1/J_2<\sqrt{2}$, whereas for stronger interaction strengths ($U\sim0.4J$) there is also large mixing with the band of higher energy states at short times when $J_1/J_2>\sqrt{2}$.

However, even for moderate interaction strengths $U=0.2 J_2$ it is still possible to achieve fidelities $F \sim 0.9$ for experimentally feasible ramp times, $J_2 T \sim 100$. This indicates that even with experimental errors in the tuning of the interaction strength, it is possible to prepare these states for sufficiently dilute and weakly interacting samples. 

\section{Many-body correlations at half-filling}
\label{sec:fermions}
There have been several previous studies on two-dimensional systems with flat bands, for example~\cite{PhysRevB.101.224514,Hub_FB,PT_FB_Lieb,PhysRevResearch.2.023136}, which utilise either mean-field theory, quantum Monte-Carlo, or dynamical mean-field theory techniques. Nevertheless,  there are still open questions here on the validity of a conventional BCS treatment in a flat band, as the interaction strength dominates the dynamics resulting in strongly correlated phases even for weak interactions.

A full quantum treatment for fermions in one-dimensional flat band systems has also been explored~\cite{PhysRevB.98.134513,PhysRevB.94.245149} with matrix product states (MPS)~\cite{SCHOLLWOCK201196}, but usually only in the isolated flat band approximation where it has been shown that the ground state can coincide with the BCS ground state~\cite{PhysRevB.94.245149}.  There are also additional questions here as to the effects on these properties in systems that have a flat band as well as dispersive bands, such as the Lieb lattice, and in particular when the interaction strengths are strong enough to mix states in different bands. However, current state of the art classical methods for quantitatively analysing these phases, such as MPS are only practical for one-dimensional systems, where it is known that Bose-Einstein condensation or superconductivity cannot occur in the ground state and instead at best the system can develop quasi-longe range order in the correlations~\cite{QP1D}. While there have been some recent numerical analysis on a minimal quasi-2D model for superconductivity~\cite{PhysRevB.100.075138}, this is an interesting opportunity to potentially use quantum simulators to improve the understanding of strongly correlated phases in 2D. 

\begin{figure}[t!]
\includegraphics[width = 8.5cm]{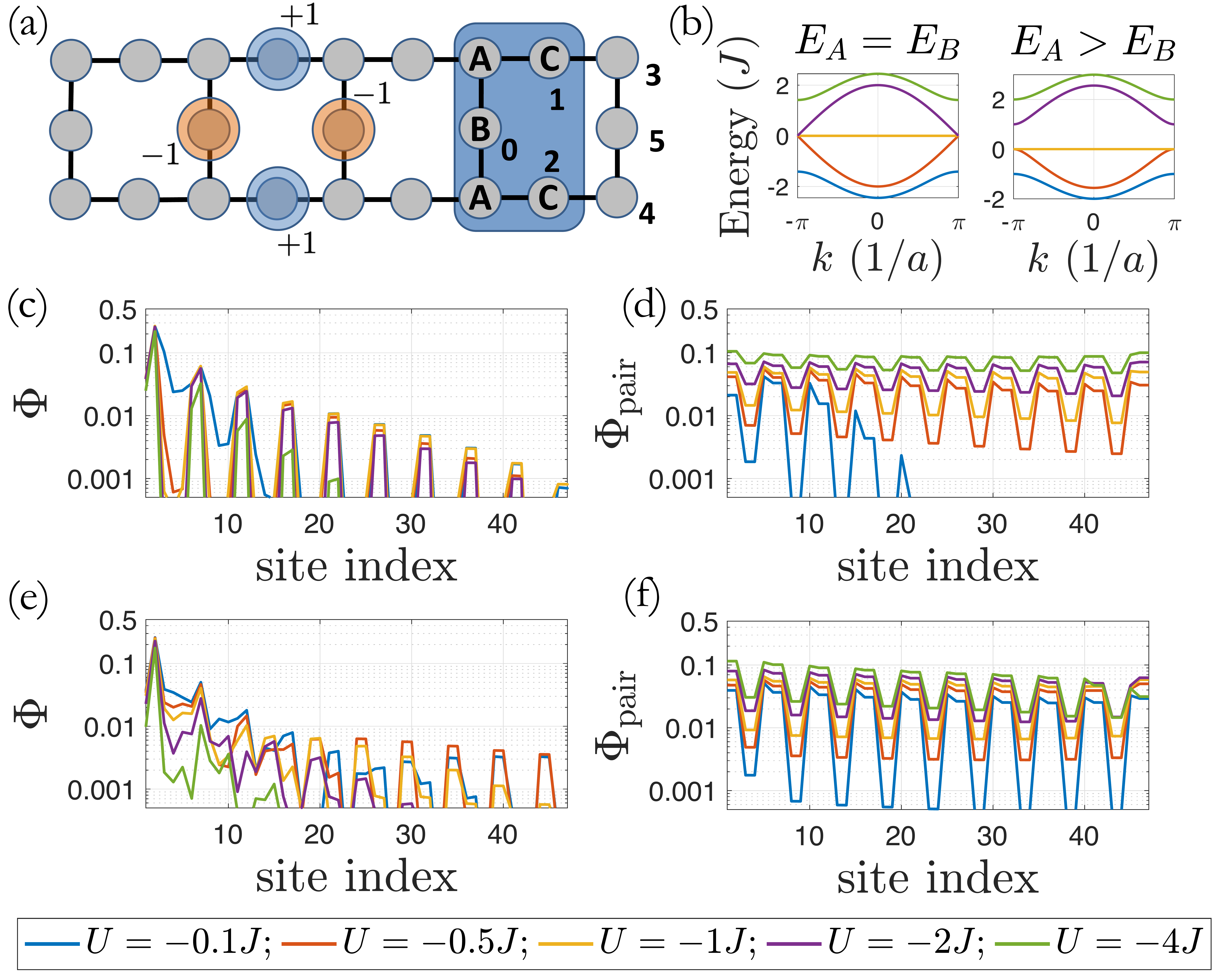}
\centering
\caption{(a) One-dimensional Lieb ladder strip with the most dominant components of the maximally localised flat band states highlighted, compare to Fig.~\ref{Wann_Lieb}(c). (b) Energy band structures for the ladder, $E_B=E_C$ for all. (c-f) Matrix product state calculation (bond dimension, $D=1024$) of the ground states at half-filling where we compare the single particle (plots (c,e), obtained using Eq.~\ref{SP_Corr}) and pair correlations (plots (d,f), obtained using Eq.~\ref{Pair_Corr}) for: (c-d) $E_A=E_B=E_C=0$ and (e-f) $E_A=J$; $E_B=E_C=0$. Note that the lattice sites in these plots are ordered according to the site index numbers in (a). System size of $20$ unit cells ($100$ sites) and $50$ spin up and $50$ spin down fermions.}
\label{Corr_Fig}
\end{figure}

In order to perform a quantitative analysis of the many-body phases produced in the Lieb lattice we focus on a one-dimensional ladder cut, shown in Fig.~\ref{Corr_Fig}, and calculate the ground state using MPS techniques. This allows us to quantitatively calculate the properties of the ground state in the strongly interacting regimes while also preserving the main qualitative features of the full Lieb lattice and providing a way to benchmark a quantum simulation of the full 2D geometry.
For this ladder system we have a similar energy band structure with a flat band and a Dirac cone (see Fig.~\ref{Corr_Fig}). In the following we consider the case of fermions at half-filling as in this regime the lower two energy bands are filled and the flat band is half filled, meaning that all excitations and dynamics around the Fermi level will be dominated by states in the flat band. Here we consider the case of uniform interaction strength on all sites of the unit cell, $U=U_A=U_{B,C}$, for simplicity, but note that we only expect small quantitative corrections to the correlation functions below as the low energy properties are dominated by the flat band Wannier basis, which have no amplitude on the A sites. This means that including an interaction imbalance on the A site should not strongly affect the behaviour we find below. Of course, for increasing interaction strength, there will be strong mixing with Wannier states in higher energy bands which do have some amplitude on the A sites, meaning that these corrections may grow as the interaction strength is increased.

We calculate the ground state for attractive interactions as the interaction strength is increased for the two band structures illustrated in Fig.~\ref{Corr_Fig}. We calculate the single particle correlations,
\begin{equation}\label{SP_Corr}
\Phi(r) = \langle a_{\uparrow,A}^{\dagger}  a_{\uparrow,r} \rangle,
\end{equation}
where we first apply the creation operator on the $A$ site in the centre of the system and then apply the destruction operator at every other site to the right in the system. We compare these to the Cooper pair correlations
\begin{equation}\label{Pair_Corr}
\Phi_{\rm pair}(r) = \langle \Delta_{B}^{\dagger}  \Delta_{r} \rangle,
\end{equation}
where
\begin{equation}
\Delta_{r}= a_{\uparrow,r} a_{\downarrow,r},
\end{equation}
and we first apply the pair creation operator to the $B$ site in the centre of the system. We plot the results in Fig.~\ref{Corr_Fig}, where we can see that for non-zero interactions the pair correlations dominate over the single particle correlations for both regimes. 

Due to the spatial distribution of the flat band eigenstates (shown in Fig.~\ref{Corr_Fig}(a)) we find that the dominant pair correlations occur between $B$ and $C$ sites, where the noticeable dips in the plots for lower interaction strengths occur when the subsequent destruction operator is applied to an $A$ site. As the interaction strength is increased these features are washed out, indicating that there is strong mixing with Wannier states associated with the higher energy bands, but intriguingly, the overall decay of the correlations seems to actually be enhanced by this mixing, whereas the single particle correlations become further suppressed. 

We can also see that in the regime $E_A=E_B=E_C$ with the Dirac cone (a-b), that the pair correlations are suppressed for weak interactions, but for strong interactions the pair correlations seem to become even larger than the case with a gap in the band structure ($E_A=J$) and for $U=-4J$ appear to remain almost constant over the entire length of the system that we have considered. Of course, with these long range correlations, finite size effects become important, making it difficult to really unequivocally determine if these correlations decay algebraically, or simply have a correlation length that is larger than the system. Nevertheless, these results are very suggestive of quasi-long range superconducting pairing, and also demonstrate that strongly mixing with dispersive energy bands, rather than destroying the novel pair dominated phases in flat bands, can actually enhance these properties.

\begin{figure}[t!]
\includegraphics[width = 8.5cm]{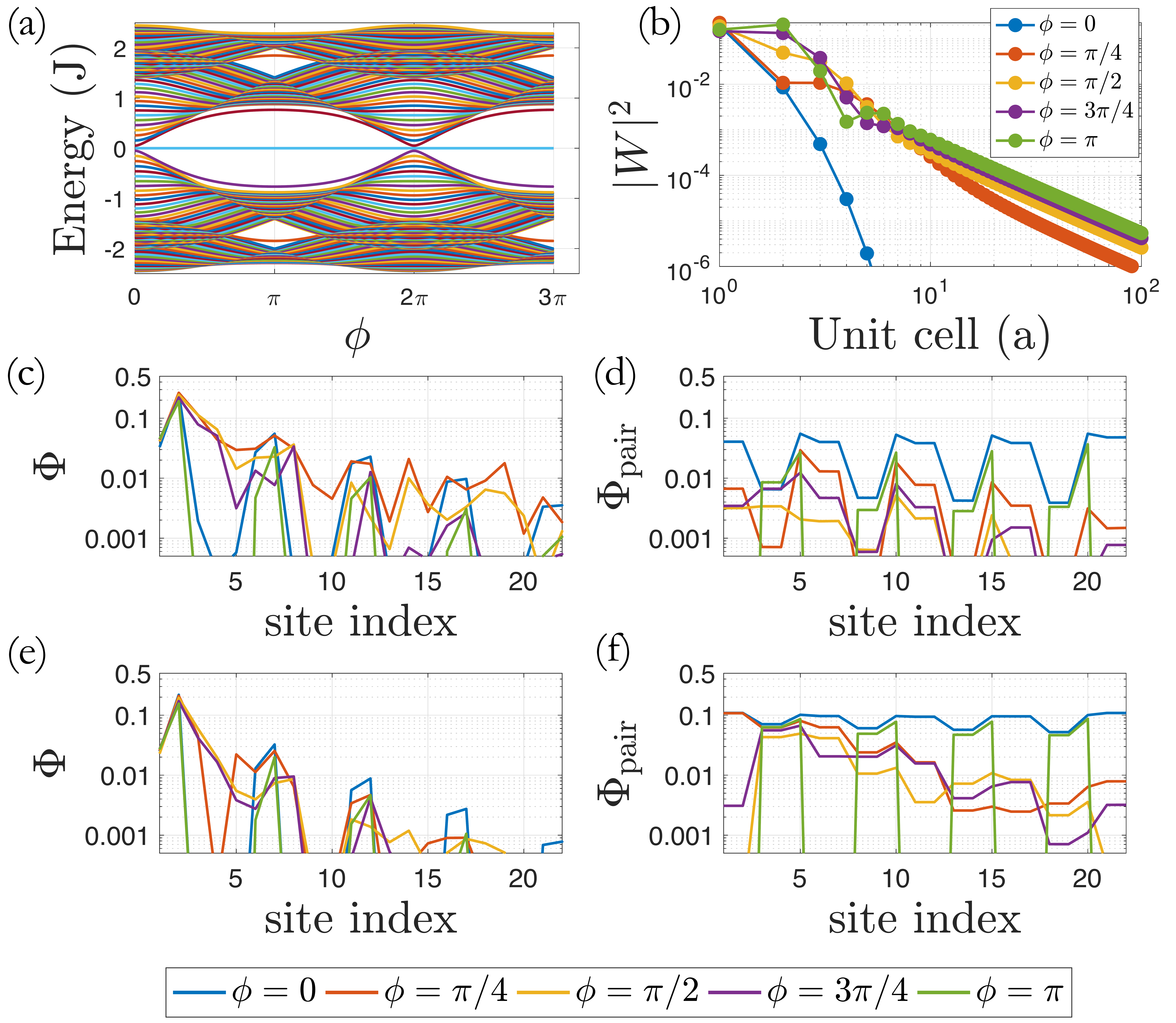}
\centering
\caption{Properties of the Lieb ladder with an applied magnetic flux, $\phi$, through each plaquette. (a) Energy eigenvalues for a Lieb ladder with $30$ unit cells and open boundary conditions as a function of $\phi$, with $E_A=E_B=E_C=0$. (b) Localised Wannier function basis for the flat band of an infinite Lieb ladder (tight-binding model) as a function of the magnetic flux. Note that each data point corresponds to the population in the unit cell. (c-f) Matrix product state calculation (bond dimension, $D=1024$) of the ground states at half-filling for different values of flux, $\phi$, where we compare the single particle (plots (c,e), obtained using Eq.~\ref{SP_Corr}) and pair correlations (plots (d,f), obtained using Eq.~\ref{Pair_Corr}) for: (c-d) $U=-0.5J$ and (e-f) $U=-4J$. Note that the lattice sites in these plots are ordered according to the site index numbers in Fig.~\ref{Corr_Fig}(a). System size of $10$ unit cells ($50$ sites) and $25$ spin up and $25$ spin down fermions.}
\label{Corr_phi_Fig}
\end{figure}

A further interesting question is how these features will change upon including a magnetic flux through the loops in each plaquette of the ladder, which we include in the Hamiltonian with a Peierls substitution~\cite{PhysRevB.89.235418}. Including these features in an experimental implementation requires modifying the above proposal, but can be achieved using Floquet engineering or laser induced tunnelling for example~\cite{Barbieroeaav7444}. In the Lieb lattice, this has been shown to introduce energy gaps between the flat band and the higher and lower dispersive bands~\cite{PhysRevB.54.R17296} which we explicitly show in Fig.~\ref{Corr_phi_Fig}(a). In Fig.~\ref{Corr_phi_Fig}(b) we calculate the single particle Wannier functions (using Eq.~\ref{equation_Wan_Vec} on an infinite tight-binding model) where we illustrate that including a magnetic flux qualitatively changes the long-range behaviour of this basis where compared to an exponential decay in the $\phi=0$ limit they develop algebraically decaying tails with degree $\sim2$ which in higher dimensions would suggest a topological Chern insulator~\cite{PhysRevLett.98.046402,Monaco:2018aa}. Additionally, we can see that at short distances these basis states become strongly extended to several nearest-neighbouring unit cells~\cite{PhysRevB.84.085124,PhysRevB.89.235418}.

We investigate how these features affect the correlations upon including strong onsite interactions by performing a similar analysis as before and we compare the single particle and pair correlations in Fig.~\ref{Corr_phi_Fig}(c-f) for a range of values for the magnetic flux, $\phi$. We compare the case for $U=-0.5J$, which is an interaction strength that is smaller than the maximum energy gap at $\phi=\pi$, to $U=-4J$ which is sufficiently strong to allow mixing between single particle states even at this maximum gap opening point. From this analysis, we can see that the overall trend in the single particle correlations (c,e) remains unaffected and these retain their exponentially decaying behaviour, indicating that the many-body effects dominate over the algebraic decay of the Wannier states even for weak interactions. However, the pair correlations also become exponentially suppressed as $\phi$ is increased, reaching a maximum suppression at $\phi \sim \pi/2$. Further increasing the flux, the correlations begin to increase again until the maximum gap opening point at $\phi=\pi$, where there is a similar algebraic like decay compared to the $\phi=0$ case, but now with a distinctly different spatial behaviour, reflecting the change in the short distance behaviour of the single particle basis states.

\section{State preparation}
\label{sec:expt}

In this section, we now consider how these many-body phases might be prepared in an experimental realisation of the Lieb lattice with an optical lattice. This preparation scheme is based on adiabatic state preparation~\cite{PhysRevA.88.012334,Simon:2011aa} and utilises the ability of the experimental systems to time-dependently vary the onsite energies, interaction strengths and tunnelling amplitudes. In order to achieve this experimentally, it requires precise knowledge of the relationship between the parameters of the optical potential and the coefficients in the effective Hubbard model which can be understood with the help of the analysis presented in the previous sections. 

\begin{figure}[t!]
{\includegraphics[width = 8.5cm]{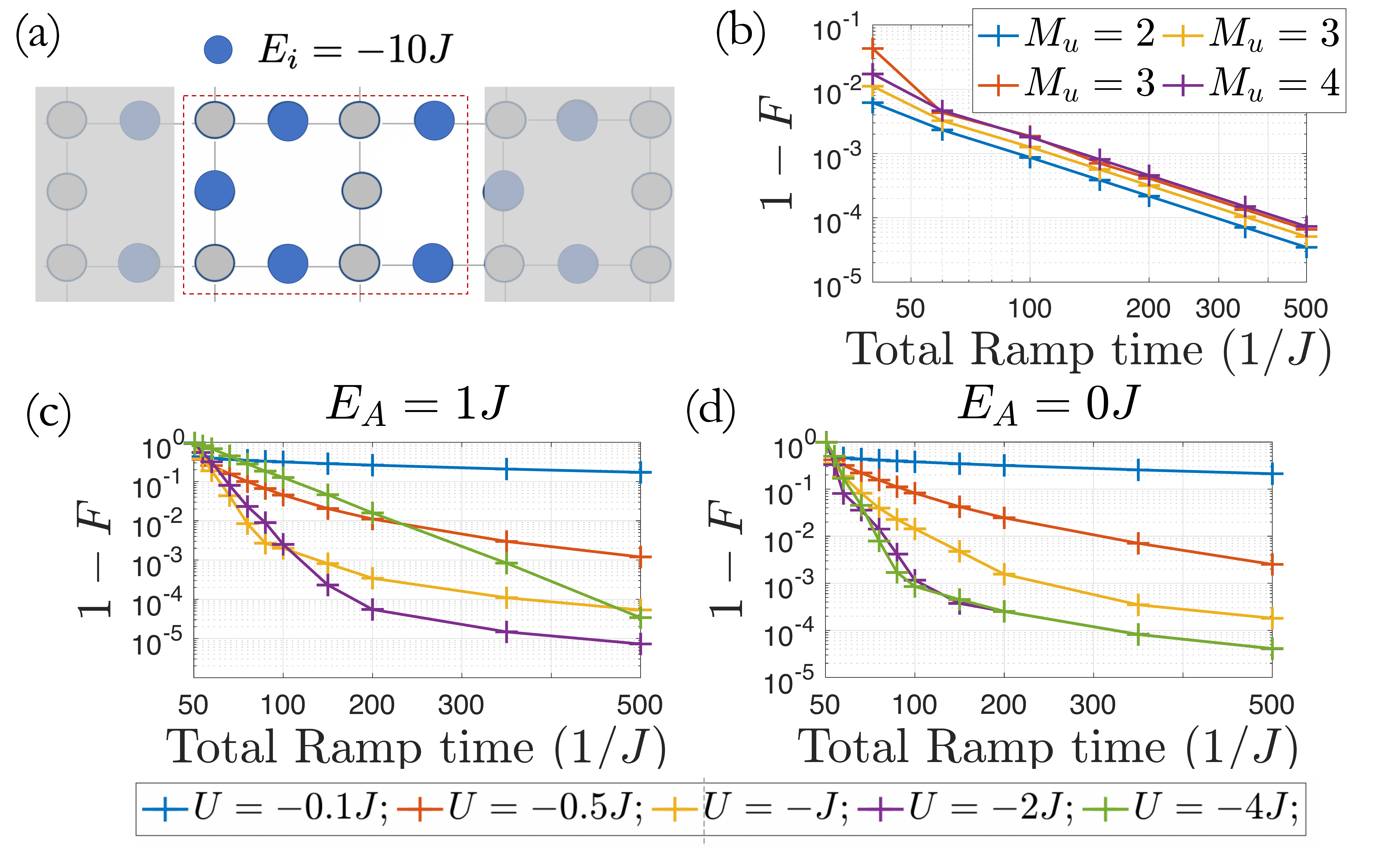}}
\centering
\caption{(a) Initial onsite energies for the adiabatic preparation process and the time dependence for the parameters. (b) The error in the fidelity, $F = |\langle \Psi_{\rm gs} | \psi_T \rangle |^2$. of the time-dependently produced state compared to the non-interacting fermion ground state as a function of the total ramp time, $T$ (linear ramp) and for different number of $5$ site unit cells (shown in Fig.~\ref{Corr_Fig}(a)). For an even number of unit cells ($M_u=2$ \& $M_u=4$) these states are at half-filling, for an odd number of unit cells ($M_u=3$) the states are at half-filling $\pm1$ atom (yellow/red). (c-d) ED results for the adiabatic preparation of the interacting ground state at half-filling for $M_u=2$, with the ramp process in Eq.~\ref{ramp_equ}, where $E_A$ is the onsite energy of the A sites, see Fig.~\ref{Corr_Fig}(a), which is constant throughout the ramp.}
\label{AdPrep_ED_Fig}
\end{figure}

We require the preparation of an initial product state where a single fermion is projected onto each of the blue sites shown in Fig.~\ref{AdPrep_ED_Fig}(a).
Our initial state in this process is the case where all tunnelling rates are zero which can be achieved with strong laser intensity creating large potential barriers and where the onsite interaction strengths are also zero, which can be tuned with a magnetic field around a Feshbach resonance~\cite{RevModPhys.82.1225}. Additionally, we require a large detuning between onsite energies with much lower energies for the sites that are initially populated, which can be achieved by varying the relative intensity of the two optical potentials creating our lattice in Eq.~\ref{equpot}. We then simultaneously time dependently ramp the tunnelling, the onsite energy and the onsite interaction strength to the desired value.

First in Fig.~\ref{AdPrep_ED_Fig}(b) we consider the case for zero interactions, with $E_A=J$; $E_B=E_C=0$, where we linearly ramp the parameters in the Hubbard model. For the non-interacting case the ground state is highly degenerate so we plot the projection of the final state onto the ground state manifold for small system sizes using exact diagonalisation. We can see that for ramp times that are achievable before decoherence effects begin to lead to errors ($TJ\sim200$) we can achieve fidelities $> 1- 1\times10^{-3}$ for the system sizes considered here.

Next we investigate the more interesting case of interacting ground states. We begin with the same initial state, but we now apply an exponential ramp for the parameters,
\begin{equation}\label{ramp_equ}
\begin{split}
E(t) &= - E_{in} \frac{e^{5(1-t/T)}-1}{e^5-1} \\
J(t) &=  J\left( 1 - \frac{e^{5(1-t/T)}-1}{e^5-1} \right) \\
U(t) &= U\left( 1 - \frac{e^{5(1-t/T)}-1}{e^5-1} \right),
\end{split}
\end{equation}
from $t=0$ to $t=T$. As the energy gap between the ground state and first excited state decreases as we ramp these parameters, with the final state being gapless in the thermodynamic limit we find it optimal to apply this exponential ramp which ensures that there is a slower change in the parameters as the gap decreases. This ensures that the adiabatic principle is not as strongly violated compared to the case for a linear ramp.  
In Fig.~\ref{AdPrep_ED_Fig}(c) we plot the projection of the final state onto the ground state using exact diagonalisation on a small system for both band structures with $E_A=J$ and $E_A=0$. We consider a range of attractive interactions where we can see that as we increase the interaction strength, the projection on the ground state improves. It is also apparent that we can achieve near perfect fidelities for experimentally achievable ramp times ($TJ\sim200$) for strong interaction strengths, which are the phases with the most dominant pair correlations, see Fig.~\ref{Corr_Fig}.

\begin{figure}[t!]
{\includegraphics[width = 8.5cm]{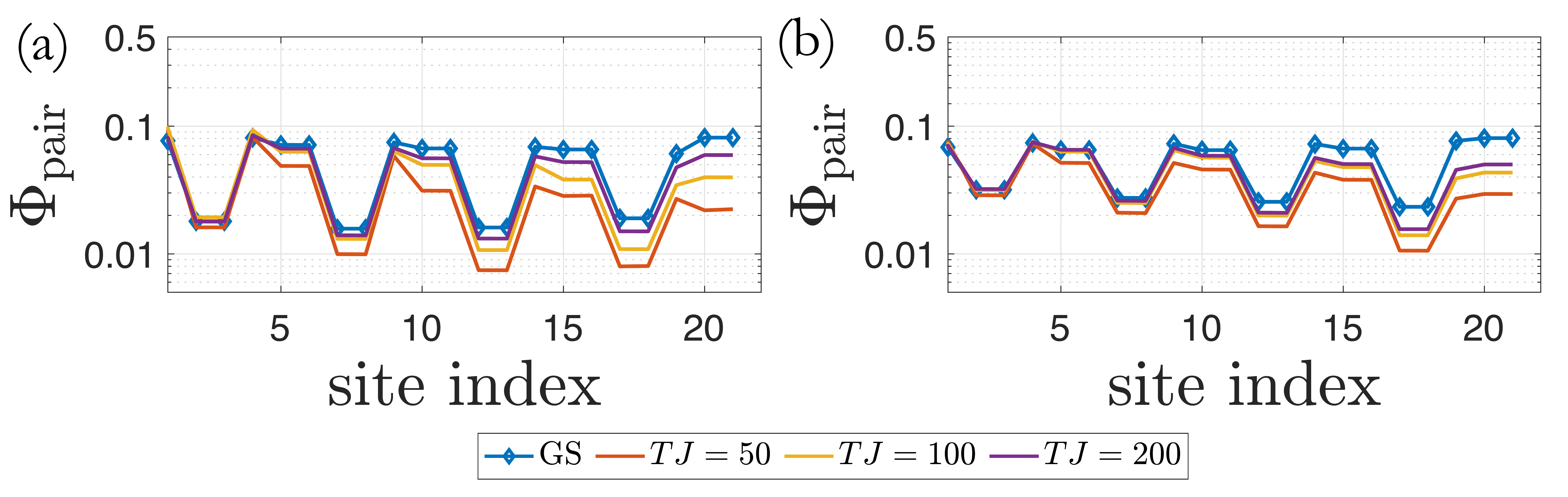}}
\centering
\caption{Comparison of the pair correlations in the half filled ground state (blue diamonds) with the time-dependently prepared state with a total ramp time, $TJ=\left(50,100,200 \right) = $ (red,yellow,purple), for $E_A=J$ (a) and $E_A=0$ (b). $U=-2J$ and system size of $10$ unit cells ($50$ sites). Calculated with a $4$th order Runge-Kutta MPS time evolution algorithm~\cite{Garcia-Ripoll:2006aa} with a timestep $J\tau=0.1$ and $D=512$. Note that the lattice sites in these plots are ordered according to the site index numbers in Fig.~\ref{Corr_Fig}(a)}
\label{AdPrep_Fig}
\end{figure}

Of course, as this analysis was carried out on small system sizes, we are only able to prepare these states so accurately due to finite size effects introducing effective energy gaps in the spectra. So, we also simulate the same ramp processes on larger system sizes using time-evolution techniques with MPS~\cite{Paeckel:2019aa,Garcia-Ripoll:2006aa}, where in Fig.~\ref{AdPrep_Fig} we compare the correlations produced through the time-dependent preparation to those for the ground state, again for both band structures, $E_A=J$ (a) and $E_A=0$ (b). 

We can see that as we increase the total ramp time, the correlations of the prepared state quickly approach those of the ground state, where for experimentally feasible ramp times ($TJ\sim100$) the deviations become too small to be resolved experimentally. Note that in all cases, the single particle correlations are exponentially suppressed. This analysis indicates that while preparing these ground states with near perfect fidelities may not be possible using adiabatic preparation schemes, due to the fact that the states are gapless in the thermodynamic limit, it is clear that we can feasibly prepare states with strong pair correlations that have a similar dependence to those of the true ground state.

This analysis touches upon an important technical consideration in these experimental architectures, namely reducing the effective temperature of the ultra-cold fermionic gas. As these atomic systems are incredibly well isolated, such that on typical experimental timescales ($TJ < 500$) they can be considered closed systems, an equivalent consideration is in reducing the entropy of the many-body system~\cite{Hilker484,PhysRevLett.120.243201,Mazurenko:2017aa}. The simulations presented in Fig.~\ref{AdPrep_Fig} indicate that, if the initial product state can be accurately prepared in the experiment (assumed to have zero entropy in the simulation) then the proposed experimental sequence can in principle be prepared with low enough entropy (and therefore effective temperature) in timescales ($TJ \sim 200$) where additional heating effects can be neglected, such as spontaneous emission~\cite{PhysRevLett.124.203201}. 

Of course there are still open questions here regarding the critical temperatures required for these pair correlations to dominate. As for typical cases with attractive pairs in optical lattices, in order to realise the proposed many-body ground states we require the temperature to be small compared to the effective dispersion relation for these pairs. In Ref.~\cite{SF_Cr} we showed that a flat band system can host stable bound pairs with a significantly enhanced pair kinetic energy which grows with interaction strength, suggesting that the critical temperature required to realise pair dominated phases may be larger compared to conventional systems and may even also grow with increasing interactions. These are important and relevant considerations which will be interesting to explore in the future.


%
%

\section{Conclusion}
\label{sec:conclusions}
We have considered an experimental realisation of the Lieb lattice, determining realistic parameters in a Hubbard model, and showing how to prepare states with weak interaction, and also phases where strong interactions give rise to  enhanced pairing correlations. By performing numerical calculations on a one-dimensional ladder system we quantitatively demonstrated that we can prepare the half filled fermion ground state in experimentally realistic timescales. Additionally we provided predictions for the ground state correlations for attractive interactions where we found that due to the flat energy band the pair correlations dominate and can be further enhanced by increasing the interaction strength such that there is strong mixing with the higher dispersive energy bands. We also found that in the strongly interacting regime, mixing between the flat band states and states close to the Dirac cone resulted in correlations that were almost constant throughout the full system that we considered.

These calculations provide a starting point for realisation of these models using parameters available in current experiments. On the theoretical side this work has further implications for understanding strongly interacting phases in flat band systems and how mixing with additional dispersive bands affects the properties of these phases. Additionally, our experimental proposal demonstrates that it is, in principle, possible to realise these phases with strong pairing correlations in current cold atom experiments. 

\begin{acknowledgements}
We thank Gerard Pelegri, Dieter Jaksch, and Hanns-Christoph N\"agerl for stimulating discussions. Work at the University of Strathclyde was supported by the EPSRC Programme Grant DesOEQ (EP/P009565/1), by AFOSR grant number FA9550-18-1-0064, and by the European Union’s Horizon 2020 research and innovation program under grant agreement No.~817482 PASQuanS. S.F. acknowledges the financial support of the Carnegie trust. Work at the University of Aveiro was funded by FCT-Portuguese Foundation for Science and Technology under the project PTDC/FIS-MAC/29291/2017, and was developed within the scope of the Portuguese Institute for Nanostructures, Nanomodelling, and Nanofabrication (i3N) Project No. UIDB/50025/2020 and UIDP/50025/2020. Results were obtained using the ARCHIE-WeSt High Performance Computer (www.archie-west.ac.uk) based at the University of Strathclyde.

All data underpinning this publication are openly available from the University of Strathclyde KnowledgeBase at https://doi.org/10.15129/69a02fc7-6454-4e1c-ba4b-5a486d1c7107.
 
 \end{acknowledgements}

\bibliography{LiebTex.bib}

\end{document}